\documentclass[12pt,preprint]{emulateapj}
\usepackage{natbib}
\usepackage{color}
\usepackage{hyperref} %----set up hyper link to citations, equations...
\hypersetup{colorlinks=true,citecolor=blue}

\def \bea {\begin{eqnarray}}
\def \ena {\end{eqnarray}}                  
\def    \simlt  {\lower.5ex\hbox{$\; \buildrel < \over \sim \;$}}
\def    \simgt  {\lower.5ex\hbox{$\; \buildrel > \over \sim \;$}}

\newcommand     \mum    {\,\mu{\rm m}}  % to use in math mode

\def	\cm		{\,{\rm {cm}}}

\def	\GHz		{\,{\rm {GHz}}}

\def	\K		{\,{\rm K}}

\def	\H		{\rm H}

\def    \Bv     	{\bf  B}

\def	\ba			{{\bf a}}

\def 	\bE		{{\bf E}}

\def	\bJ		{{\bf J}}

\def	\gas		{\rm {gas}}

\def	\d			{\rm d}

\def    \ext    	{\rm {ext}}
\def    \pol    	{\rm {pol}}

\def	\gra		{\rm {gra}}
\def	\sil	 	 {\rm {sil}}
\def	\carb		{\rm {carb}}

\def    \Planck		{{\it Planck}~}

\def	\PAH		{\rm {PAH}}

\def	\obs		{\rm {obs}}
\def	\mod		{\rm {mod}}

\begin{document}
\shorttitle{Polarized Spinning Dust Emission and UV polarization}
\shortauthors{Hoang, Lazarian, \& Martin}
\title{Constraint on the Polarization of Electric Dipole Emission from Spinning Dust}

\author{Thiem Hoang\altaffilmark{1}, A. Lazarian\altaffilmark{2}, and P. G. Martin\altaffilmark{1}}
\altaffiltext{1}
{Canadian Institute for Theoretical Astrophysics, University of Toronto,
60 St. George Street, Toronto, ON M5S 3H8, Canada
}
\altaffiltext{2}
{ Department of Astronomy, University of Wisconsin-Madison, Madison, WI 53705, USA
}

\begin{abstract}
\Planck results have revealed that the electric dipole emission from polycyclic aromatic hydrocarbons (PAHs) is the most reliable explanation for anomalous microwave emission that interferes with cosmic microwave background (CMB) radiation experiments. The emerging question is to what extent this emission component contaminates to the polarized CMB radiation. We present constraints on polarized dust emission for the model of grain size distribution and grain alignment that best fits to observed extinction and polarization curves. Two stars with a prominent polarization feature at $\lambda=2175$ \AA, HD 197770 and HD 147933-4, are chosen for our study. For HD 197770, we find that the model with aligned silicate grains plus weakly aligned PAHs can successfully reproduce the 2175 \AA~polarization feature; whereas, for HD 147933-4, we find that the alignment of only silicate grains can account for that feature. The alignment function of PAHs for the best-fit model to the HD 197770 data is employed to constrain polarized spinning dust emission. We find that the degree of polarization of spinning dust emission is about $1.6\%$ at frequency $\nu \approx 3\GHz$ and declines to below $0.9\%$ for $\nu>20\GHz$. We also predict the degree of polarization of thermal dust emission at $353\GHz$ to be $P_{\rm em}\approx 11\%$ and $14\%$ for the lines of sight to the HD 197770 and HD 147933-4 stars, respectively.
\end{abstract}

\section{\label{sec61}Introduction}

Cosmic Microwave Background (CMB) experiments (see \citealt{Bouchet:1999p4616}; \citealt{Tegmark:2000p5597}; \citealt{Efstathiou:2003p4793}; \citealt{Bennett:2003p4582}) are of great importance for studying the early universe and its subsequent expansion. Precision cosmology with {\it Wilkinson Microwave  Anisotropy Probe} 
({\it WMAP}) and \Planck requires a good model of the microwave foreground emission to allow the reliable subtraction of foreground contamination from the CMB radiation. 

In the 10--60 GHz frequency range, electric dipole emission (\citealt{1998ApJ...508..157D}) from rapidly spinning tiny dust grains (mostly polycyclic aromatic hydrocarbons, hereafter PAHs; \citealt{1984A&A...137L...5L}; \citealt{1992A&A...253..498R}) is an important component of Galactic foregrounds that dominates the CMB signal (see \citealt{Collaboration:2013vx}). In the last several years, significant progress has been made in understanding spinning dust in terms of both theory (\citealt{2009MNRAS.395.1055A};\citealt{2010A&A...509A..12Y}; \citealt{2010ApJ...715.1462H}; \citealt{2011ApJ...741...87H}; \citealt{Silsbee:2011p5567}; \citealt{2013AdAst2013E...2A}; see \citealt{Hoang:2012bb} for a review) and observation (\citealt{Dickinson:2009p607}; \citealt{2011ApJ...734....4K}; \citealt{2012ApJ...754...94T}). In particular, {\it Planck} results have confirmed spinning dust emission as the most reliable source of anomalous microwave emission (AME) (\citealt{PlanckCollaboration:2011p515}). {\it Planck} is poised to release interesting results on the CMB polarization; however, the question to what extent the spinning dust emission contaminates to the polarized CMB signal remains unclear.

The degree of polarization of spinning dust emission depends on the alignment efficiency of PAHs in the interstellar magnetic field. \cite{2000ApJ...536L..15L} (hereafter LD00) suggested that PAHs can be aligned via resonance paramagnetic relaxation--a mechanism which extends the classical Davis-Greenstein (\citealt{1951ApJ...114..206D}) mechanism for very fast rotating grains for which the Barnett magnetization  arising from fast rotation of grains cannot be neglected. They predicted the polarization of spinning dust typically $\leq 1\%$ for frequency $\nu>20$ GHz.

Observational studies (\citealt{2006ApJ...645L.141B}; \citealt{2009ApJ...697.1187M}; \citealt{2011MNRAS.418L..35D}; \citealt{2011MNRAS.418..888M}) showed that the upper limit for the AME polarization is between $1-5\%$. In addition, an upper limit of $1\%$ for the AME polarization is reported in various media (see \citealt{LopezCaraballo:2011p508}; \citealt{RubinoMartin:2012ji}). Since \Planck and other CMB experiments provide extremely precise measurements of polarization, a reliable prediction for the spinning dust polarization is useful for separation of polarized Galactic foreground components from the CMB.

The problem of grain alignment, especially for PAHs\footnote{The accepted theory of grain alignment for large grains is that based on radiative torques (\citealt{1976Ap&SS..43..291D}; \citealt{1996ApJ...470..551D}; \citealt{1997ApJ...480..633D}; \citealt{2007MNRAS.378..910L}; \citealt{2008ApJ...676L..25L}; \citealt{2008MNRAS.388..117H}; \citealt{2009ApJ...695.1457H}; \citealt{2009ApJ...697.1316H}) but the radiative torques are negligibly small for PAHs. This encourages us to explore the possibilities provided by other alignment mechanisms.} is complicated in general (see \citealt{2007JQSRT.106..225L} for a review), but one can derive constraints on grain alignment observationally (see \citealt{2007EAS....23..165M}). An important attempt to obtain observational constraints on grain alignment was carried out by \cite{1995ApJ...444..293K}. The authors applied maximum entropy method to infer the mass distribution of aligned grains through fitting theoretical polarization curves to observational data. They discovered that interstellar silicate grains of size $a \ge 0.05\mum$ are aligned, whereas smaller grains, including PAHs, are weakly aligned. \cite{Draine:2009p3780} performed simultaneous fitting to the typical extinction and polarization curves of the diffuse interstellar medium (ISM) and came to the similar conclusion as in \cite{1995ApJ...444..293K} that small ($a<0.05\mum$) grains are weakly aligned and large ($a>0.1\mum$) grains are efficiently aligned. 

Since the tiny dust grains that radiate spinning dust emission are likely the same as those that produce the ultraviolet (UV) extinction bump at $\lambda=2175$\AA,\footnote{Such a UV bump is believed to arise from the electronic transition $\pi-\pi^{*}$ in the $sp^{2}$-bonded carbon sheets of small carbonaceous grains (see \citealt{1989IAUS..135..313D}; \citealt{2007ApJ...657..810D}).} a good way to search for the alignment of PAHs is through its imprint on the UV polarization. While the UV extinction bump is ubiquitous in the ISM, the UV polarization bump is rarely seen, except for two stars HD 197770 and HD147933-4, which show a prominent $2175$\AA~polarization feature (\citealt{1992ApJ...385L..53C}; \citealt{1993ApJ...403..722W}; \citealt{1997ApJ...478..395W}). In addition to the UV polarization bump, HD 197770 exhibits an excess UV polarization that cannot be accounted for with the typical ISM polarization curve--Serkowski law (\citealt{Serkowski:1973p6351}) as do some stars with the peak wavelength of polarization curve $\lambda_{\max}<0.55\mum$ (\citealt{1995ApJ...445..947C}; \citealt{1999ApJ...510..905M}). Such an excess UV polarization may be due to the enhanced alignment of small silicate grains by paramagnetic relaxation or radiative torques (Hoang et al. 2013, submitted).  

A number of studies have suggested that the UV polarization bump can arise from aligned, small graphite grains (\citealt{1988ApJ...333..848D}; \citealt{1993ApJ...403..722W}; \citealt{1997ApJ...478..395W}); however, a detailed study quantifying the alignment efficiency of small graphite grains is not yet available.

The main goal of this paper is to find the degrees of alignment of interstellar grains including PAHs that reproduces the 2175\AA~ polarization bump seen in HD 197770 and HD 147933-4 and employ the inferred degrees of alignment to predict the degree of polarization of spinning dust emission. The paper is structured as follows.

In \S \ref{sec:optical} we calculate extinction cross section and polarization cross section for silicate and carbonaceous oblate spheroidal grains. In \S \ref{sec:fitUV}, we describe a procedure to derive grain size distributions and degree of grain alignment by fitting theoretical predictions to observed extinction and polarization curves and present the obtained results. In \S \ref{sec:polspdust} polarized spinning dust emission is calculated using the degree of grain alignment obtained for the best-fit model. Discussion and summary are presented in \S \ref{sec:discus} and \S \ref{sec:summ}, respectively.

\section{Optical properties of dust grains}\label{sec:optical}

\subsection{Extinction and Polarization Cross Section}
Interstellar dust grains are widely known to induce the extinction and polarization of starlight due the absorption and scattering of light out of the line of sight by dust grains.

Let us consider an oblate spheroidal grain with the symmetry axis $\ba_{1}$ having an effective size $a$, which is the size of an equivalent sphere of the same volume as the grain. A perfectly polarized electromagnetic wave with the electric field vector $\bE$ is assumed to propagate along the line of sight. Let $C_{\ext}(\bE\perp \ba_{1})$ and $C_{\ext}(\bE\| \ba_{1})$ be the extinction of radiation by the grain for the cases in which $\bE$ is parallel and perpendicular to $\ba_{1}$, respectively. For the sake of simplification, we denote these extinction cross sections by $C_{\perp}$ and $C_{\|}$. 

For the general case in which $\bE$ makes an angle $\theta$ with $\ba_{1}$, the extinction cross section becomes 
\bea
C_{\ext}=\cos^{2}\theta C_{\|}+\sin^{2}\theta C_{\perp}.\label{eq:Cext0}
\ena

Since the original starlight is unpolarized, one can compute the total extinction cross section for a randomly oriented grain by integrating Equation (\ref{eq:Cext0}) over the isotropic distribution of $\theta$, i.e., $f_{\rm iso} d\theta \sim \sin\theta d\theta$. As a result,
\bea
C_{\ext}=\frac{1}{3}\left(2C_{\perp}+C_{\|}\right).\label{eq:Cext}
\ena

The polarization coefficient for oblate spheroidal grains is equal to
\bea
C_{\pol}=\left[C_{\ext}(\bE \| \ba_{1})-C_{\ext}(\bE \perp \ba_{1})\right].
\ena

The dust extinction and polarization cross section can be represented through the extinction and polarization efficiency, which are respectively defined by:
\bea
Q_{\ext}(a,\lambda)=\frac{C_{\ext}(a,\lambda)}{\pi a^{2}},\\
Q_{\pol}(a,\lambda)=\frac{C_{\pol}(a,\lambda)}{\pi a^{2}}.
\ena

\subsection{Silicate grains}
For silicate grains, we employ the publicly available DDSCAT code (\citealt{2012arXiv1202.3424D}) to compute the extinction cross section $C_{\ext}^{\sil}(a,\lambda)$ using the dielectric functions of amorphous silicate from \cite{2003ApJ...598.1026D}.

\begin{figure}
\includegraphics[width=0.45\textwidth]{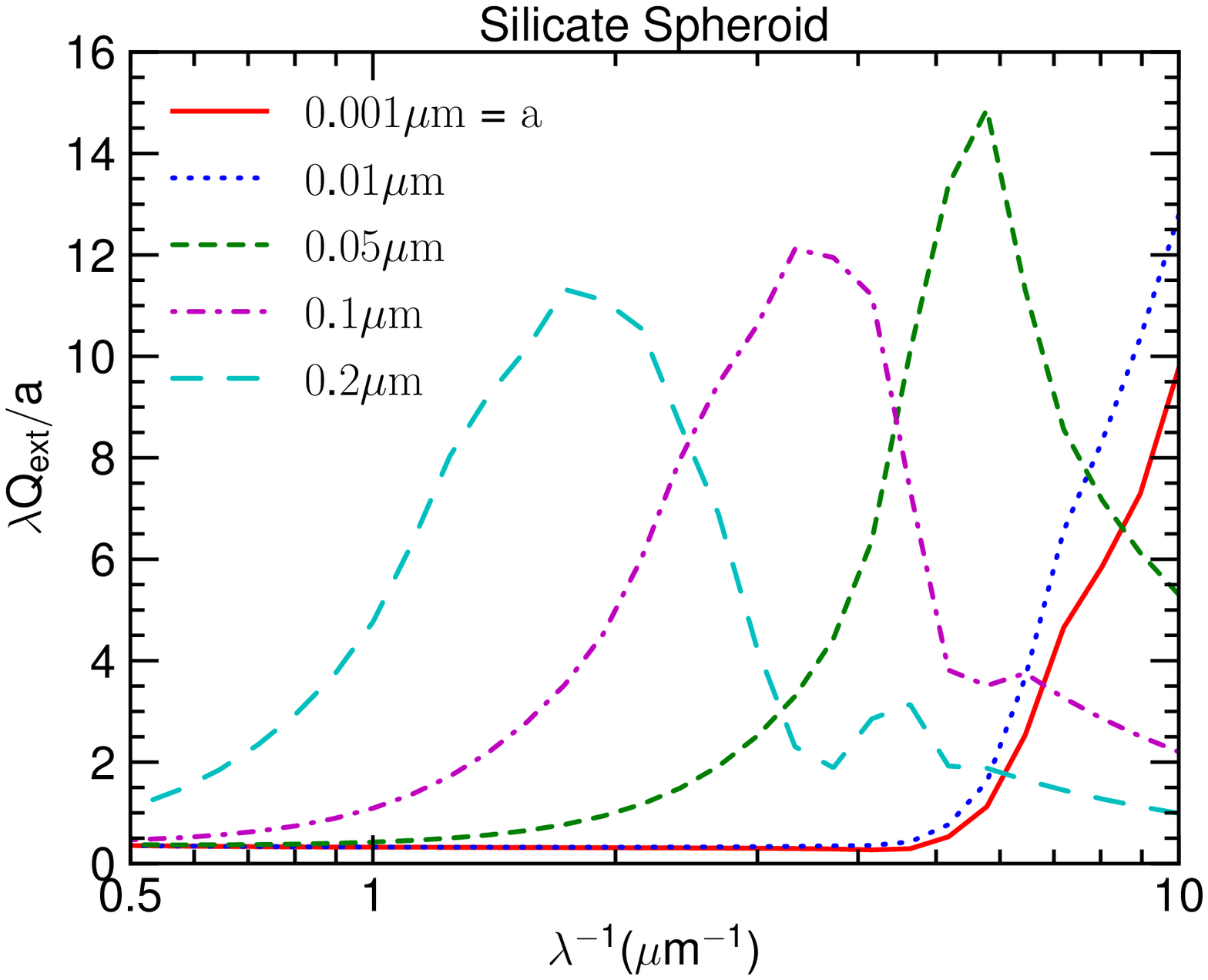}
\includegraphics[width=0.45\textwidth]{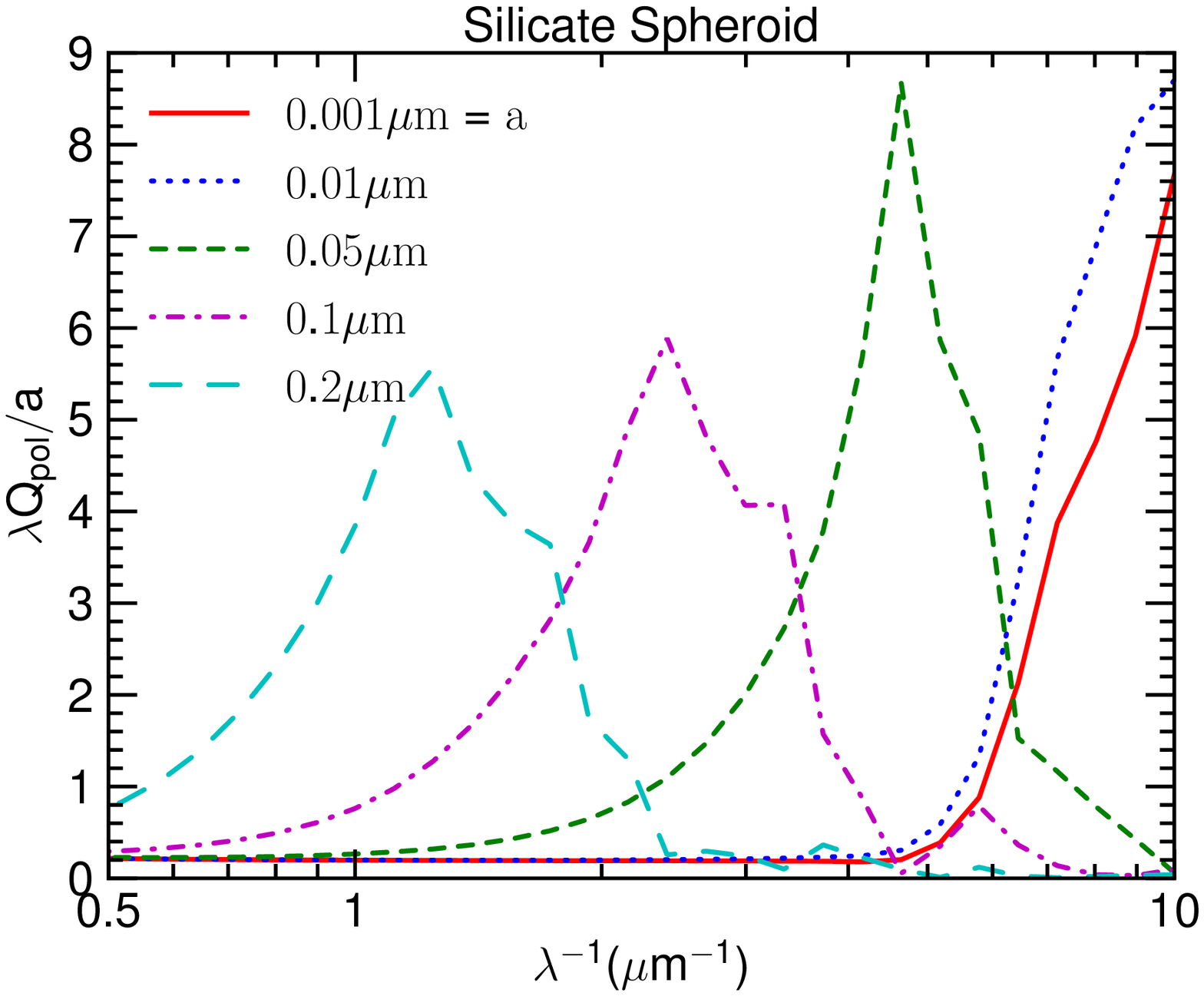}
\caption{Extinction efficiency $\lambda Q_{\ext}/a$ (upper) and polarization efficiency $\lambda Q_{\pol}/a$ (lower) of silicate grains as functions of $1/\lambda$ for the different grain sizes.}
\label{fig:Qext_waveinv}
\end{figure}

Figure \ref{fig:Qext_waveinv} shows $\lambda Q_{\ext}/a$ and $\lambda Q_{\pol}/a$ as functions of $\lambda^{-1}$ for a silicate spheroidal grain with axial ratio $r=2$ and for the different grain sizes.

\subsection{Carbonaceous grains}
Carbonaceous grains consist of a fraction $\zeta_{\PAH}$ of PAHs and $1-\zeta_{\PAH}$ of graphite grains with
\bea
\zeta_{\PAH}(a)=(1-q_{\gra})\times \min\left[1,(a_{\zeta}/a)^{3}\right],
\ena
where $a_{\zeta}=50$\AA, $q_{\gra}=0.01$ are assumed as in \cite{2007ApJ...657..810D}.

For graphite grains, which are an anisotropic material, we adopt the dielectric functions from \cite{2003ApJ...598.1026D} and compute $C_{\ext}^{\gra}(a,\lambda)$ using DDSCAT. We consider two cases in which $\bE$ are parallel and perpendicular to the grain optical axis (c-axis) with the corresponding the dielectric function $\epsilon_{\|}$ and $\epsilon_{\perp}$. Using the $\frac{1}{3}-\frac{2}{3}$ approximation (i.e., 1/3 of graphite grains have $\epsilon=\epsilon_{\|}$ and 2/3 of them have $\epsilon=\epsilon_{\perp}$), one can obtain the extinction cross section $C_{\ext}^{\gra}=\left(C_{\ext}(\bE\| c)+2C_{\ext}(\bE\perp c)\right)/3$ for a given grain orientation. Using Equation (\ref{eq:Cext}) we obtain $C_{\ext}$ for randomly oriented grains.

For PAHs, the extinction $C_{\ext}^{\PAH}(\lambda)$ is computed as in Equations (5)-(10) in \cite{2001ApJ...554..778L}. The polarization cross section of PAHs is taken to be the same as graphite grains of the same size. It has been shown in \cite{1992A&A...266..513V} that the absorption properties (and then polarization properties) of PAHs are similar to those of bulky graphite grains when the dielectric constant of bulky graphite is adopted for PAHs.

\subsection{Extinction and Polarization Curves}\label{sec:theory}
The extinction induced by randomly oriented grains in units of magnitude is defined by
\bea
A({\lambda})&=&2.5{\rm log}_{10}\left(\frac{F_{\lambda}^{\obs}}{F_{\lambda}^{\star}}\right)=1.086\tau_{\lambda},\nonumber\\
&=&1.086\int \sum_{j=\carb,\sil}\sum_{i=0}^{N_{a}-1} C_{\ext}^{j}(a_{i})
n^{j}(a_{i})dz,\label{eq:Aext}
\ena
where $F_{\lambda}^{\star}$ is the intrinsic flux from the star, $F_{\lambda}^{\obs}=F_{\lambda}^{\star}e^{-\tau{_\lambda}}$ is the observed flux, $\tau_{\lambda}$ is the optical depth, and the integration is performed along the line of sight.

To find the polarization by aligned grains, let us define an observer's reference system in which the line of sight is directed along the $z-$axis, and $x-$ and $y-$ axes constitute the sky plane. The polarization of starlight arising from the dichroic extinction by aligned grains in a cell of $dz$ is computed as
\bea
dp({\lambda})=\frac{d\tau_{x}-d\tau_{y}}{2}=\sum_{i=0}^{N_{a}-1}\frac{1}{2}\left(C_{x}-C_{y}\right)
n(a_i)dz,
\label{eq:dplam}
\ena
where $N_{a}$ is the number of size bin, $n(a_{i})\equiv (dn/da)da$ is the number of grains of size $a_{i}$, $C_{x}$ and $C_{y}$ are the grain cross section along the $x-$ and $y-$ axes, respectively.

For the case of perfect internal alignment of grain symmetry axis $\ba_{1}$ with angular momentum $\bJ$, by transforming the grain coordinate system to the observer's reference system and taking corresponding weights, we obtain
\bea
C_{x}&=&C_{\perp}-\frac{C_{\pol}}{2}\sin^{2}\beta,\\
C_{y}&=&C_{\perp}-\frac{C_{\pol}}{2}(2\cos^{2}\beta\cos^{2}\gamma+\sin^{2}\beta\sin^{2}\gamma),
\ena
where $\gamma$ is the angle between the magnetic field $\Bv$ and the sky plane and $\beta$ is the angle between $\bJ$ and $\Bv$.

The polarization efficiency then becomes
\bea
C_{x}-C_{y}=C_{\pol}\frac{\left(3\cos^{2}\beta-1\right)}{2}\cos^{2}\gamma.\label{eq:Cpol}
\ena
Taking the average of $C_{x}-C_{y}$ over the distribution of the alignment angle $\beta$, it yields
\bea
C_{x}-C_{y}=C_{\pol} Q_{J}\cos^{2}\gamma,\label{eq:Cx-Cy}
\ena
where $Q_{J}=\langle G_{J}\rangle$ is the ensemble average of $G_{J}=\left(3\cos^{2}\beta-1\right)/2$ that describes the alignment of $\bJ$ and $\Bv$.

When the internal alignment is not perfect, following the similar procedure, we obtain
\bea
C_{x}-C_{y}=C_{\pol}\langle 	G_{J}G_{X} \rangle \cos^{2}\gamma \equiv
C_{\pol}R\cos^{2}\gamma,\label{eq:Cpol}
\ena
where $G_{X}=\left(3\cos^{2}\theta-1\right)/2$ with $\theta$ being the angle between $\ba_{1}$ and $\bJ$, and $R=\langle G_{J}G_{X}\rangle$ is the Rayleigh reduction factor (see also \citealt{1999MNRAS.305..615R}). The degree of internal alignment is described by $Q_{X}=\langle G_{X}\rangle$.

For a perpendicular magnetic field, i.e., $\Bv$ lies on the sky plane, Equation (\ref{eq:Cpol}) simply becomes $C_{x}-C_{y}=C_{\pol}R$. For an arbitrary magnetic field geometry, let $f=R\cos^{2}\gamma$ be the effective degree of grain alignment, which is a function of grain size $a$. Thus, in the following, $f(a)$ is referred as the alignment function.

Plugging this above equation into Equation (\ref{eq:dplam}) and integrating along the line of sight, we obtain
\bea
p({\lambda})=\int \sum_{j=\carb,\sil}\sum_{i=0}^{N_{a}-1} \frac{1}{2}C_{\pol}^{j}(a_{i})f^{j}(a_{i})
n^{j}(a_i)dz,\label{eq:Plam}
\ena
where $f^{j}(a_i)$ is the effective degree of grain alignment for the grain specie $j$ of size $a_i$. 

It is more convenient to represent the extinction (polarization) through the extinction (polarization) cross section. Hence, the above equations can be rewritten as
\bea
A({\lambda})&=&1.086~\sigma_{\ext}(\lambda)\times N_{\H},\label{eq:sigext}\\
p({\lambda})&=&\sigma_{\pol}(\lambda)\times N_{\H},\label{eq:sigpol}
\ena
where $N_{\H}(\cm^{-2})$ is the column density, and $\sigma_{\ext}$ and $\sigma_{\pol}$ in units of $\cm^{2} \H^{-1}$ are the dust extinction cross section and polarization cross section, respectively.

\section{Grain size distribution and alignment functions constrained by observations}\label{sec:fitUV}
\subsection{Nonlinear Chi-square Fitting}
In this section, we find grain size distributions and alignment functions by fitting theoretical predictions to the observational data for the HD 197770 and HD 147933-4 stars. The parameters for these stars, including the optical depth $\tau_{V}$, ratio of visual to selective extinction $R_{V}=A_{V}/E_{B-V}$, peak polarization $p_{\max}$, peak wavelength $\lambda_{\max}$, and polarization efficiency $p_{\max}/A(\lambda_{\max})$, are listed in Table \ref{tab:HD} (see also \citealt{1997ApJ...478..395W}).

\begin{table}[h]
\caption{Optical extinction and polarization parameters}\label{tab:HD}
\begin{tabular}{l l  l l l l} \hline\hline\\
Star HD & $R_{V}$ & $\tau_{V}$ & $\lambda_{\max}(\mu\rm m)$  & $p_{\max}(\%)$ & $p_{\max}/A(\lambda_{\max})$\cr\\
\hline\cr
197770 &  3.1 & 1.66 & $0.511\pm 0.001$ & $3.90 \pm 0.01$ & 1.95 $\%$/mag\cr
147933-4  & 4.3 & 1.86 & $0.683\pm 0.003$ & $2.72 \pm 0.01$ & $1.67$ $\%$/mag \cr
\cr
\hline
\end{tabular}
\end{table}

We adopt the mixture dust model consisting of amorphous silicate grains, graphite grains and PAHs (see \citealt{2001ApJ...548..296W} ; \citealt{2007ApJ...657..810D}). 
Since there is no observational evidence for a particular shape of interstellar grains, here we assume that both silicate and carbonaceous grains are oblate spheroidal as in previous studies (e.g., \cite{1995ApJ...444..293K}; \citealt{2006ApJ...652.1318D}; \citealt{Draine:2009p3780}). These studies also show that a wide range of the grain axial ratio $r$ can successfully reproduce the observational data. Moreover, our principal interest is focused on PAHs, a reasonable assumption for the axial ratio of PAHs is necessary to infer a reliable constraint for their degree of alignment. Indeed, a disklike shape with radius $R$ and height $L$ is usually assumed for PAHs with $a \le 6$\AA~(see e.g. \citealt{1998ApJ...508..157D}). The grain volume is $V= 4\pi a^3/3= 4\pi r^{-1} b^3/3 = \pi R^2 L$ with $L=3.35$\AA~for planar PAHs. For the smallest PAH with $a =3.56$\AA~that is the most abundant, we obtain $R^2=4a^3/3L \approx 25$ or $R \approx 5$\AA. If we approximate the disklike shape as an oblate spheroid such that the length of the major semiaxis is equal to the disk radius (i.e., $b=R$), then the axial ratio $r=(4/3) b^{3}/(L R^2) \approx 4R/3L \approx 2$. As a result, we conservatively assume the axial ratio $r=2$ for all grains throughout the paper.

In the present paper, we assume that graphite grains are randomly oriented, whereas silicate grains and PAH can be aligned. Such an assumption allows us to obtain an upper limit on the alignment efficiency of PAHs since aligned graphite grains would contribute some polarization to the $2175$\AA~feature. As in \cite{Draine:2009p3780}, we simultaneously fit theoretical predictions to the observed extinction and polarization curves. 

Following \cite{1995ApJ...444..293K}, we find the grain size distribution and alignment function by minimizing a  objective function $\chi^{2}$, which is given by
\bea
\chi^{2}=\chi_{\ext}^{2}+\chi_{\pol}^{2}+\chi_{\rm con}^{2},\label{eq:chisq}
\ena
where
\bea
\chi_{\ext}^{2}=w_{\ext}\sum_{i=0}^{N_{\lambda}-1}\left[A_{\mod}(\lambda_{i})-A_{\obs}(\lambda_{i})\right]^{2},\\
\chi_{\pol}^{2}=w_{\pol}\sum_{i=0}^{N_{\lambda}-1}\left[p_{\mod}(\lambda_{i})-p_{\obs}(\lambda_{i})\right]^{2},
\ena
with $w_{\ext}$ and $w_{\pol}$ being the fitting weights for the extinction and polarization, respectively, and the last term $\chi_{\rm con}^{2}$ contains the additional constraints of the fitting model. Here, the summation is performed over $N_{\lambda}$ wavelength bins. We consider $N_{a}=64$ size bins from $a=3.56$\AA~to $1\mum$ ~and $N_{\lambda}=64$ from $\lambda=0.15$ to $2.5\mum$. Both grain size and wavelength grids have their logarithms equally spaced.

The ``observed" extinction $A_{\obs}(\lambda)$ is calculated using the extinction law (\citealt{1989ApJ...345..245C}; \citealt{1994ApJ...422..158O}) for the measured values $R_{V}$ of the two stars (see Table \ref{tab:HD}), and the ``observed" polarization $p_{\obs}(\lambda)$ is obtained by interpolating the observational data from \cite{1997ApJ...478..395W} for $N_{\lambda}$ wavelength bins. The extinction $A_{\mod}(\lambda)$ and polarization $p_{\mod}(\lambda)$ are given by Equations (\ref{eq:Aext}) and (\ref{eq:Plam}). 

Following \cite{2006ApJ...652.1318D}, we introduce a number of constraints for grain size distribution $dn/da$ and alignment function $f(a)$. The most important constraint for $f(a)$ is given by the polarization efficiency $p_{\max}/A(\lambda_{\max})$. For the case of the maximum $p_{\max}/A(\lambda_{\max})=3\% {\rm mag}^{-1}$ (\citealt{Serkowski:1975p6681}), the conditions for grain alignment are expected to be optimal such that the alignment of big grains can be perfect, and the magnetic field is regular and perpendicular to the line of sight. Thus, $f(a=a_{\max})=1$ has been taken for this optimal case (\citealt{Draine:2009p3780}; Hoang et al. 2013). For HD 197770 and HD 147933-4 having lower $p_{\max}/A(\lambda_{\max})$, the constraint for $f(a)$ should be adjusted such that $f(a=a_{\max})=(1/3)p_{\max}/A(\lambda_{\max})$. Moreover, for relatively large ($a>0.05\mum$) grains that are aligned by radiative torques (see \citealt{2007JQSRT.106..225L}), we expect the degree of alignment increasing with $a$. In particular, our detailed calculations in Hoang et al. (2013) show that $f(a)$ of small grains that are aligned by Davis-Greenstein paramagnetic relaxation tends to increase monotonically with $a$ whereas $f(a)$ of very small ($a<50$\AA) grains aligned by resonance relaxation does not. Thus, a constraint for the monotonic increase of $f(a)$ versus $a$ is introduced for grains larger than $50$\AA. Other constraints include the non-smoothness of $dn/da$ and $f(a)$ (see \citealt{2006ApJ...652.1318D}).

The fitting procedure is started with an initial size distribution $n(a)$ that best reproduces the observational data for the diffuse ISM, which corresponds to model 3 in \cite{Draine:2009p3780}. By doing so, we implicitly assume that dust properties are similar throughout the ISM and the difference in the polarization of starlight is mainly due to the efficiency of grain alignment, which depends on environment conditions along the line of sight, e.g., radiation field, magnetic fields and gas density. For silicate grains, we take the alignment function for the ISM from \cite{Draine:2009p3780} as an initial alignment function.  For PAHs, we adopt a physically motivated initial alignment function that is directly computed in Hoang et al. 2013. This initial alignment function peaks at $a\sim 10$\AA.

The nonlinear least square fitting is carried out using the Monte Carlo method. Basically, for each size bin, we generate $N_{\rm rand}$ random samples in the range $[-\zeta,\zeta]$    from a uniform distribution for $f(a)$ and $n(a)$, $\alpha_{f}$ and $\alpha_{n}$, respectively. The new values of $f$ and $n$ are given by $\tilde{f}=(\alpha_{f}+1)f(a)$ and $\tilde{n}=(\alpha_{n}+1)n(a)$. Then we calculate $A_{\mod}$ and $p_{\mod}$ for the new values $\tilde{f}$ and $\tilde{n}$ using Equations (\ref{eq:Aext}) and (\ref{eq:Plam}). The values of $\chi^{2}$ obtained from Equation (\ref{eq:chisq}) are used to find the minimum $\chi^{2}$. The range $[-\zeta,\zeta]$ of the uniform distribution is adjusted after each iterative step. Initially $\zeta=0.5$ is assumed, which allows more room for the random sampling, and when the convergence is close (i.e., the variation of $\chi^{2}$ is small) $\zeta$ is decreased to $\zeta=0.1$.

The above process is repeated until convergence criterion is satisfied. Here, we employ the convergence criterion which is based on the decrease of $\chi^{2}$ after one step: $\epsilon=(\chi^{2}(n,f)-\chi^{2}(\tilde{n},\tilde{f}))/\chi^{2}(n,f)$. If $\epsilon\le \epsilon_{0}$ with $\epsilon_{0}$ sufficiently small, then the convergence is said to be achieved. For $\epsilon_{0}=10^{-3}$, we found that the convergence is achieved after $32$ steps for HD 197770. The convergence is hard to achieve for HD 147933-4 due to the rather low UV polarization relative to the peak polarization, so we stop the iterative process after 43 steps.

\subsection{Results}
\begin{figure}
\includegraphics[width=0.45\textwidth]{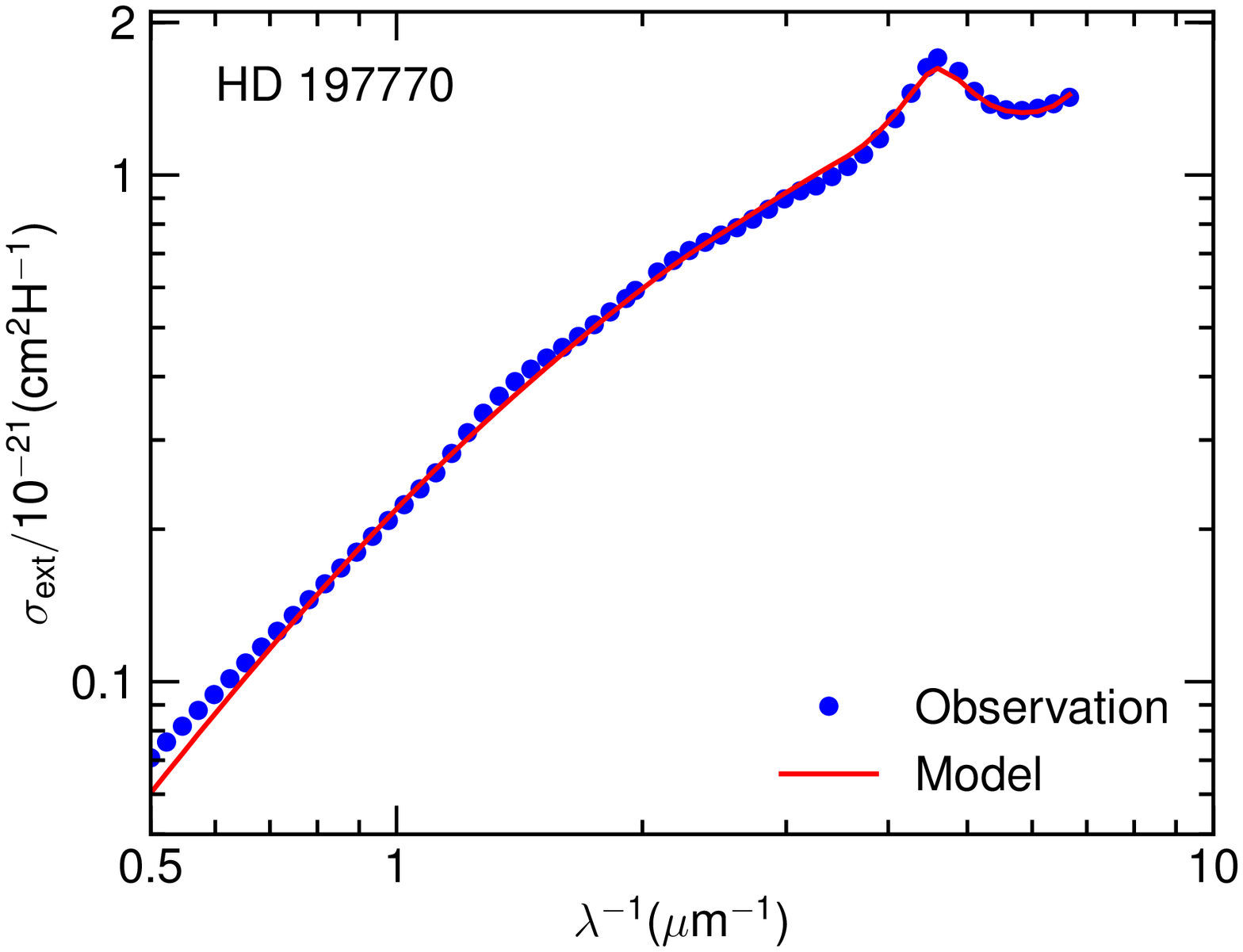}
\includegraphics[width=0.45\textwidth]{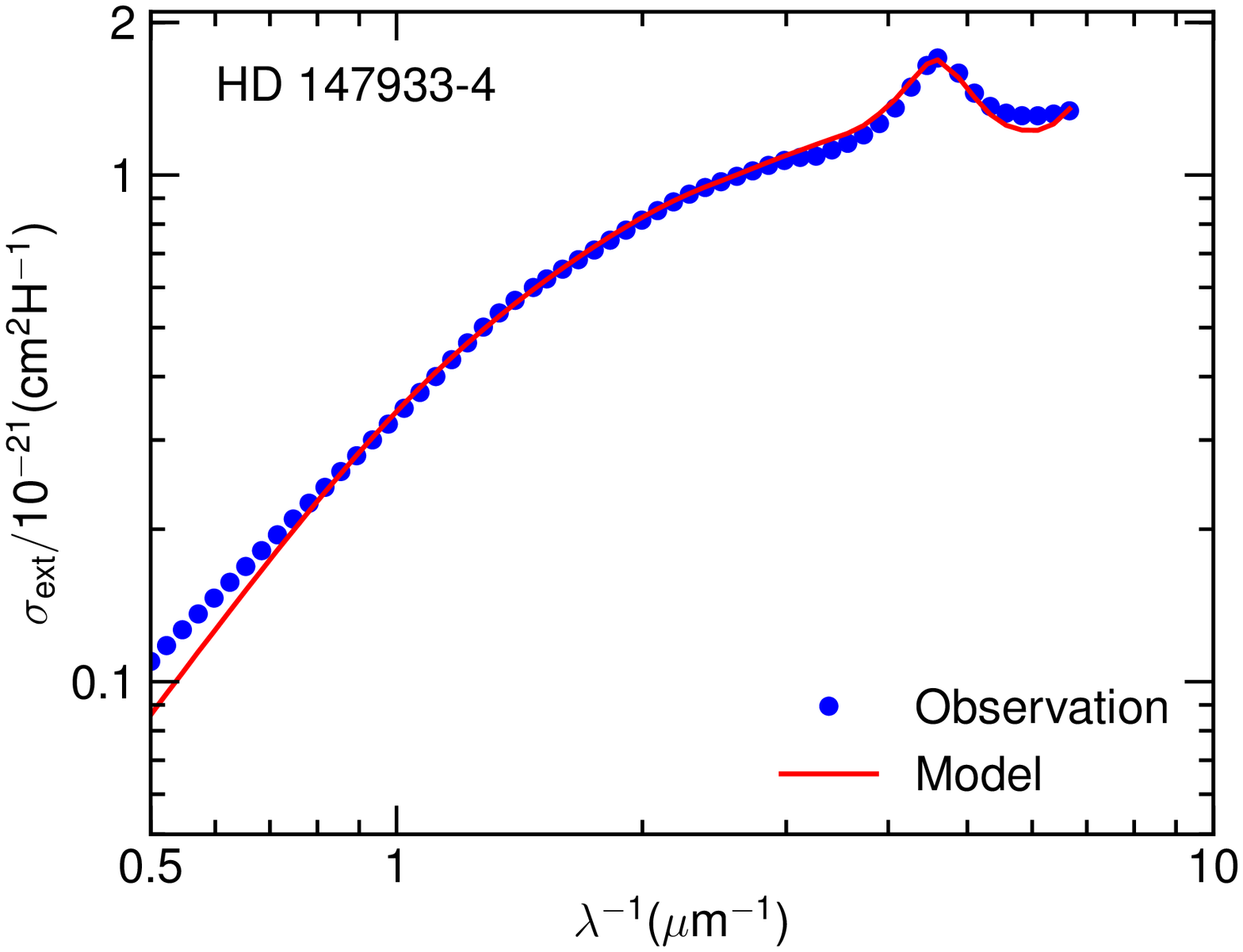}
\caption{Observed extinction curve (filled ellipses) versus model for HD 197770 and HD 147933-4. An excellent fit achieved for both stars.}
\label{fig:extcurve}
\end{figure}

Figure \ref{fig:extcurve} shows the extinction curves from our best-fit model compared to the ``observed" data for the HD 197770 and HD 147933-4 stars. As shown the model can successfully reproduce the observed extinction curves for both stars. 

\begin{figure}
\includegraphics[width=0.45\textwidth]{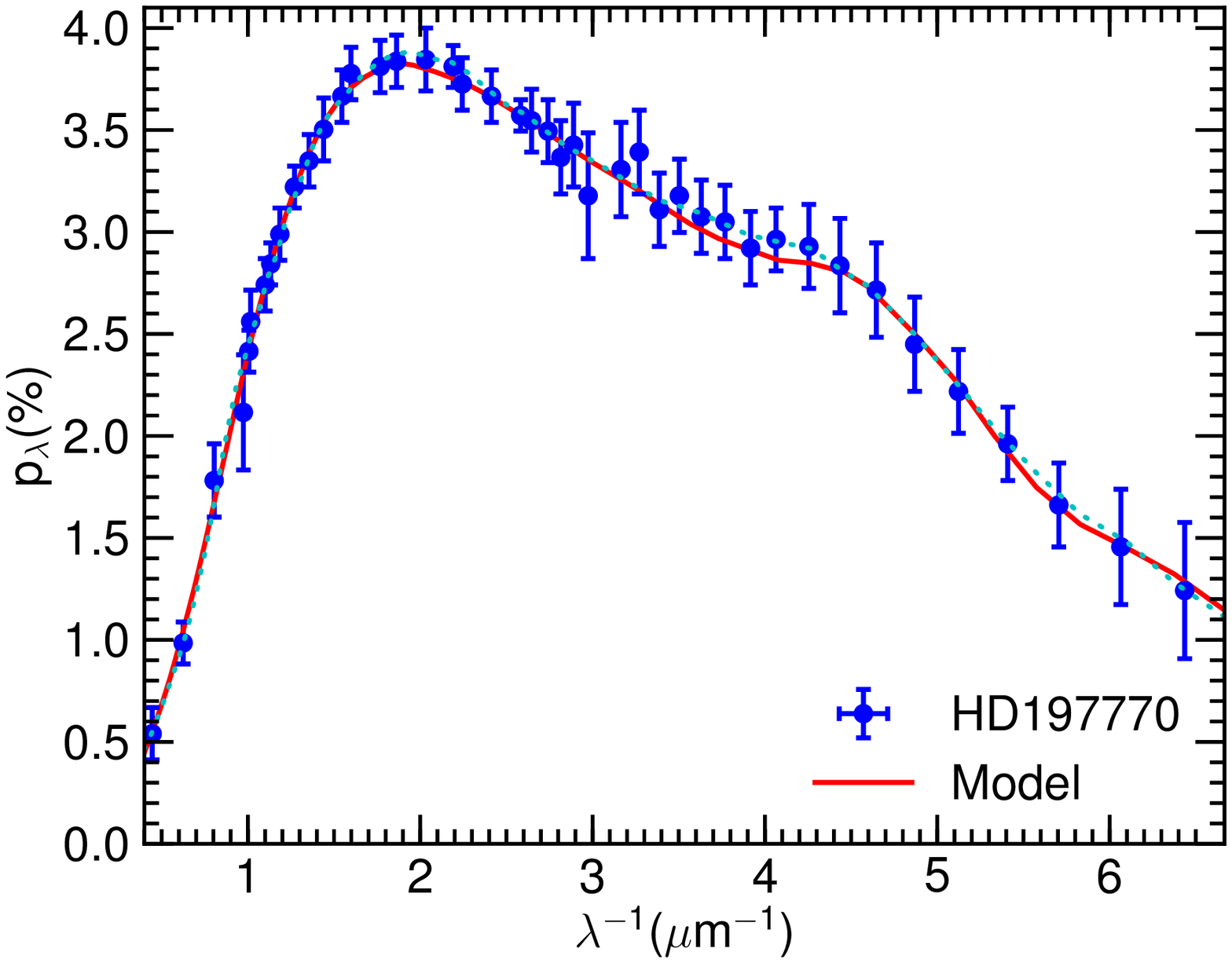}
\includegraphics[width=0.45\textwidth]{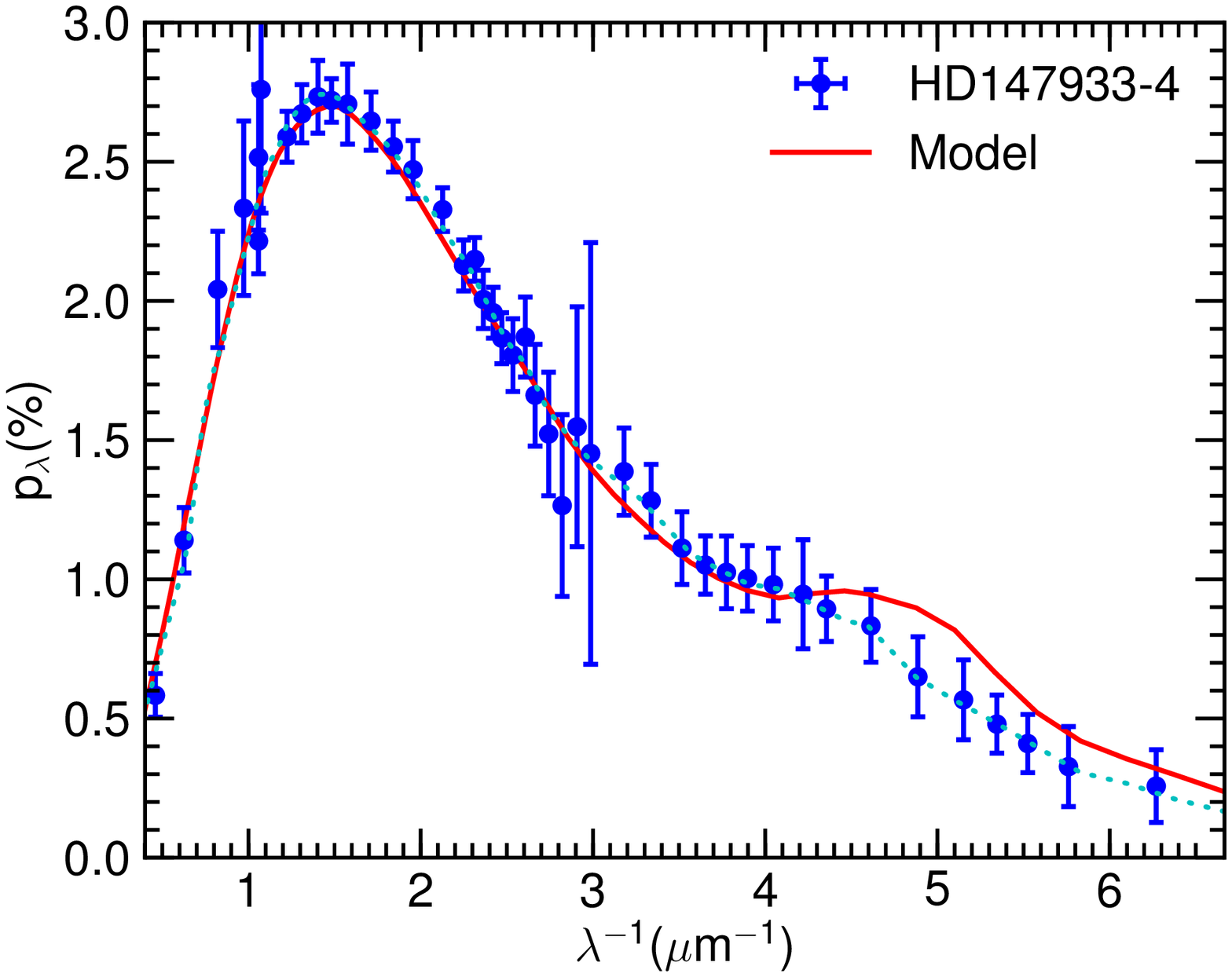}
\caption{Polarization curves for our model (solid lines) versus the observed polarization data (filled circles) for HD 197770 (upper) and HD 147933-4 (lower). Dotted lines show the observational data interpolated to our wavelength bins used for model fitting. The 3$\sigma$ error bars are shown. An excellent fit achieved for HD 1977770.}
\label{fig:polfit}
\end{figure}

Figure \ref{fig:polfit} shows the polarization curves from our best-fit model versus the ``observed" data from \cite{1997ApJ...478..395W}. Here the error bars of $3\sigma$ are shown. For HD 197770, it can be seen that our model provides an excellent fit to the observed polarization, successfully reproducing both the $2175$ \AA~polarization feature and excess UV polarization. 

The $2175$ \AA~polarization feature in HD 147933-4 can also be reproduced, but the model appears to overestimate the UV polarization for $\lambda^{-1}>5\mum^{-1}$ (see the lower panel in Figure \ref{fig:polfit}). Such a poor fit to the low UV polarization present in the line of sight with large $\lambda_{\max}$ arises from the fact that the UV polarization signal is dominated by relatively large grains, which produce the optical and infrared polarization. Thus, any improvement in the fit to the UV polarization comes with the expense of a poorer fit to the optical and infrared polarization (see also \citealt{1995ApJ...444..293K}).

To see quantitatively the goodness of our best-fit model, we calculate the chi-square test statistics: 
\bea
{X}_{\ext}^{2}=\sum_{i=0}^{N_{\lambda}-1}\frac{\left[A_{\mod}(\lambda_{i})-A_{\obs}(\lambda_{i})\right]^{2}}{A_{\obs}(\lambda_{i})},\\
X_{\pol}^{2}=\sum_{i=0}^{N_{\lambda}-1}\frac{\left[p_{\mod}(\lambda_{i})-p_{\obs}(\lambda_{i})\right]^{2}}{p_{\obs}(\lambda_{i})},
\ena
using the best-fit parameters. We obtain $X^{2}_{\ext}=0.16$ and $X^{2}_{\pol}=0.05$ for HD 197770. The similar values $X^{2}_{\ext}=0.27$ and $X^{2}_{\pol}=0.59$ for HD 147933-4. Thus, the best-fit model appears to be good for both stars.

\begin{figure}
\includegraphics[width=0.45\textwidth]{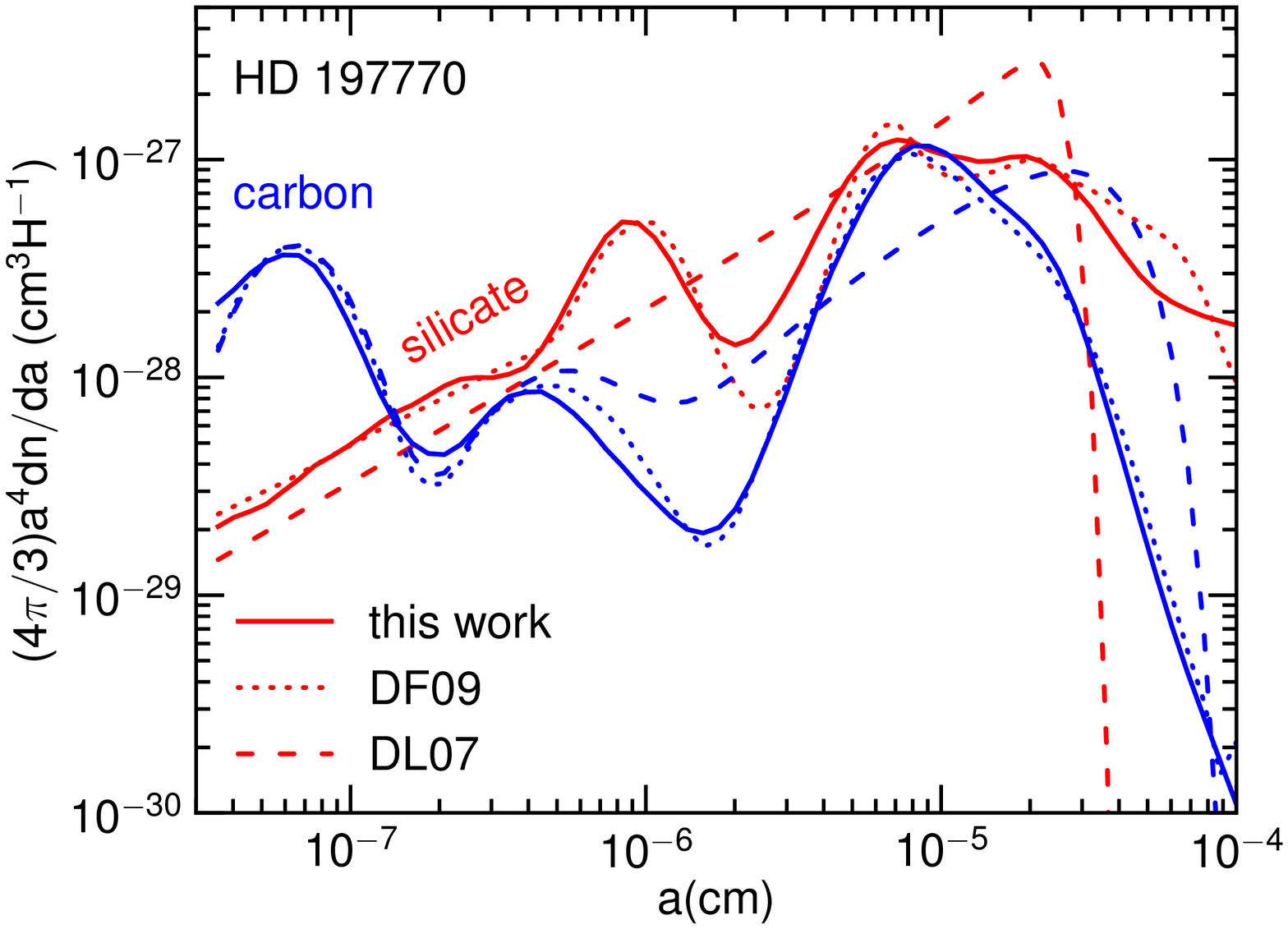}
\includegraphics[width=0.45\textwidth]{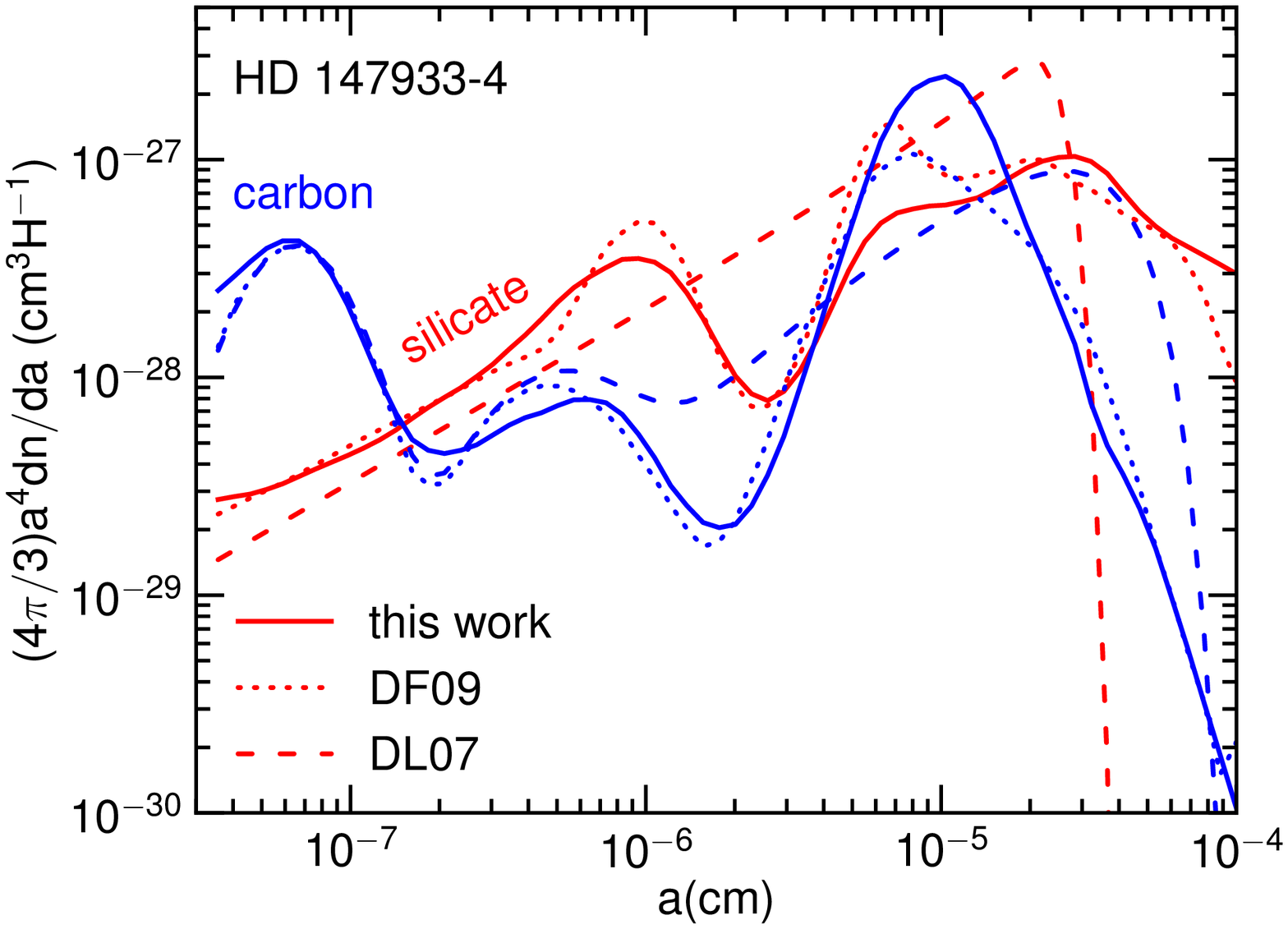}
\caption{Mass distribution of silicate grains and carbonaceous grains for the best-fit model (solid lines) for the HD 197770 (upper) and HD 147933-4 (lower) stars. Dotted (DF09) and dashed (DL07) lines show the grain mass distributions from \cite{Draine:2009p3780} and \cite{2007ApJ...657..810D}.}
\label{fig:mafit}
\end{figure}

Figure \ref{fig:mafit} shows the mass distributions of silicate and carbonaceous grains for the best-fit model for HD 197770 (upper) and HD 1479333-4 (lower). The distribution functions (DL07 and DF09) from \cite{2007ApJ...657..810D} and \cite{Draine:2009p3780} are shown for comparison.

For HD 197770, it can be seen that the mass of PAHs in our model is similar to DL07 and DF09. The mass distribution of graphite grains is nearly the same as DF09 and distinct from DL07. For silicate grains, the mass distribution peaks at $a\sim 0.009\mum, 0.06\mum$ and $0.2\mum$. Compared to DF09, our model requires a slight increase of silicate mass in the range $a=[0.012\mum-0.05\mum]$ to reproduce the excess UV continuum polarization seen in HD 197770. 

For HD 147933-4, the mass distribution of silicate grains peaks at a larger size ($a\approx 0.25\mum$) than for HD 197770, whereas the mass of small grains ($a<0.2\mum$) is subtantially reduced. The peak in the mass of ($\sim 0.1\mum$) graphite grains is increased compared to DF09. Such a decrease in the mass of small grains and increase in the mass of large grains is necessary to reproduce the higher value $R_{V}$ along the sightline to HD 147933-4 ($R_{V}=4.3$) compared to the typical value ($R_{V}=3.1$) along most of the sightlines of the diffuse ISM.

\begin{figure}
\includegraphics[width=0.45\textwidth]{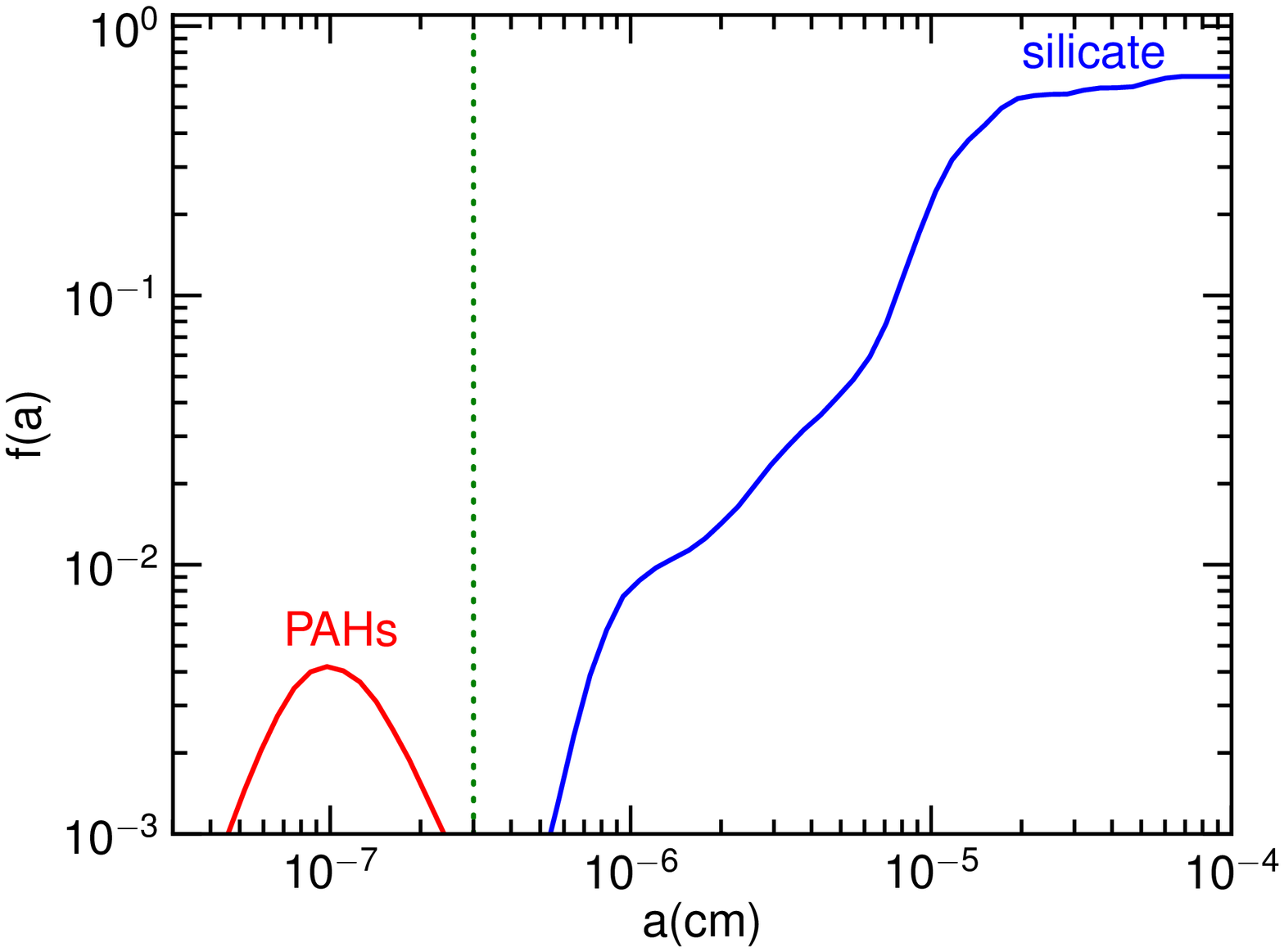}
\includegraphics[width=0.45\textwidth]{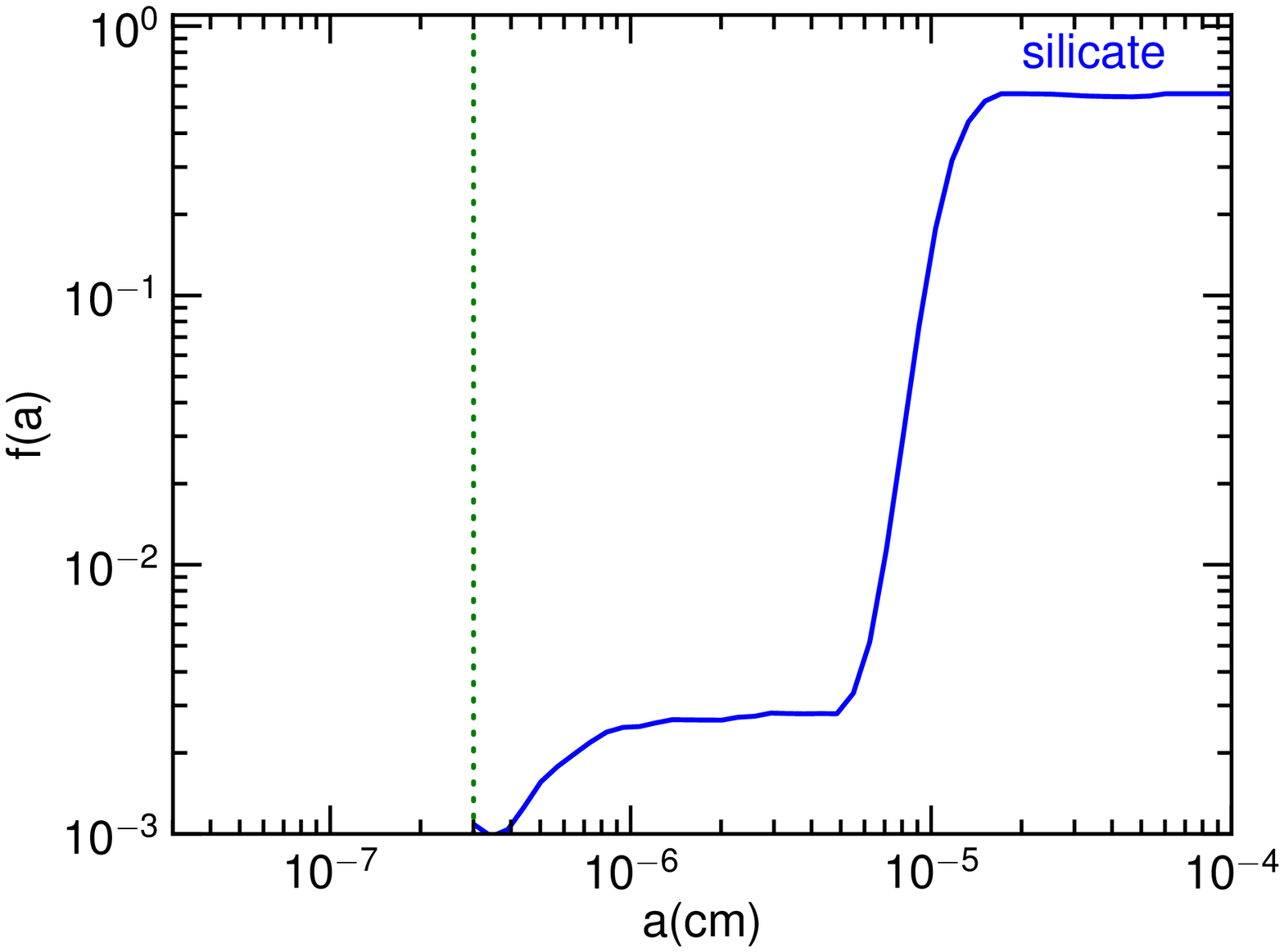}
\caption{{Upper panel:} effective degree of alignment as a function of grain size for silicate grains (blue line) and carbonaceous grains (red line) obtained from the fitting for HD 197770. Dotted line shows $a=3\times 10^{-7}\cm$ from which the polarization by aligned small silicate grains becomes negligible. {Lower panel:} similar to upper panel but for HD 147933-4. The alignment of silicate grains is shown only since the presence of PAH alignment is unnecessary (see the text).}
\label{fig:Rfit}
\end{figure}

Figure \ref{fig:Rfit} (upper) shows the alignment function $f(a)$ that reproduces the polarization curve in Figure \ref{fig:polfit} (upper). We can see that the alignment of silicate grains increases monotonically with the increasing $a$ and becomes nearly constant with $f\sim 0.6$ for $a>0.2\mum$. The degree of alignment declines rapidly for $a<0.01\mum$. The alignment of $a<0.003\mum$ (i.e. 30\AA) silicate grains is not shown because the polarization by tiny silicate grains containing minor dust mass is negligible. In particular, one can see from the figure that the best-fit model corresponds to a peaky alignment function of PAHs with $f_{\rm peak}\approx 0.004$ at $a\approx 10$\AA.\footnote{This value is a factor of 2 larger than the rough estimate $\Delta p/\Delta \tau \sim 0.0017$ in \cite{1997ApJ...478..395W}.} It indicates that a rather low alignment degree of PAHs could be sufficient to reproduce the UV polarization feature. 

For HD 147933-4 (see Figure \ref{fig:Rfit}, lower), we find that the best-fit model requires no alignment of PAHs (i.e., the alignment of only silicate grains can reproduce the observed polarization). The degree of alignment $f\sim 0.55$ for $a>0.15\mum$ silicate grains. Moreover, the alignment of silicate grains decline steeply for $a<0.1\mum$ and becomes negligible ($f<0.003$) for $a<0.05\mum$ (small grains). It turns out that small silicate grains have negligibly small contribution to the UV polarization, which is different from HD 197770. It is noted that the sharp decline in alignment of small ($a<0.1\mum$) silicate grains (i.e., the polarization is dominated by larger silicate grains) is required to account for the large value of $\lambda_{\max}$ seen in HD 147933-4. 

\section{Polarization of spinning dust emission}\label{sec:polspdust}
\subsection{Degree of grain alignment for spinning dust emission}
To calculate polarized spinning dust emissivity, it is important to understand the relation between the degree of alignment of PAHs that is responsible for the polarization in microwave emission and the degree of alignment of PAHs that results in starlight polarization by extinction (i.e., Rayleigh reduction factor).

As discussed in \cite{2007JQSRT.106..225L}, these two measures are expected to be different. Indeed, even spherical PAHs aligned by paramagnetic mechanism can produce polarized microwave emission. Naturally, the polarization by UV absorption or polarized infrared emission is not expected from the spherical PAHs. Although we do not really believe that most of PAHs are spherical, this illustrates the problem that we face comparing different observational consequences of alignment. For instance, fast wobbling and flipping grains are also expected to deliver substantially reduced degrees of polarization in terms of UV absorption. However, this just decreases the polarization of microwave emission by a factor of unity\footnote{One can say that for the microwave polarization from PAHs the alignment in terms of the PAH angular momenta is important, while for the UV absorption polarization, it is the alignment in terms of grain axes that is essential. The two alignments are different (see \citealt{1997ApJ...484..230L};\citealt{2008MNRAS.388..117H}; \citealt{2011ApJ...741...87H}).}. 

In the following, to find the upper limit of polarized spinning dust, we employ the degree of alignment of angular momentum with the magnetic field, $Q_{J}$. Thus, from the inferred alignment function $R(a)$, we can derive $Q_{J}$ using the following relationship (see Section \ref{sec:theory})
\bea
R(a)\equiv \langle G_{X}G_{J}\rangle =Q_{J}\times Q_{X}(1+f_{\rm corr}),\label{eq:Ra}
\ena
where $Q_{X}$ describes the degree of alignment of grain axes with the angular momentum and $f_{\rm corr}$ describes the correlation between the internal and external alignment. The case $f_{\rm corr}=0$ indicates that the internal alignment is independent from the external alignment. Our calculations show that $f_{\rm corr}\sim 0.6$ for $a=10$\AA~grains.

\begin{figure}
\includegraphics[width=0.5\textwidth]{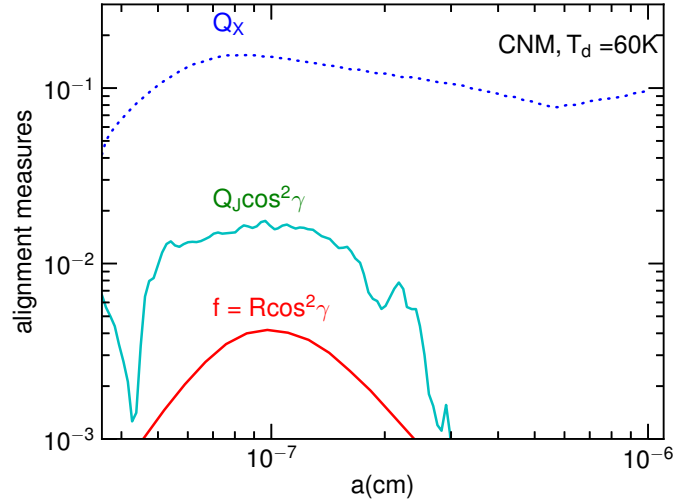}
\caption{Alignment measures as functions of grain size $a$: effective degree of alignment $f=R\cos^{2}\gamma$, degree of alignment of grain axes with the angular momentum $Q_{X}$, and effective degree of alignment of angular momentum with the magnetic field $Q_{J}\cos^{2}\gamma$. $Q_{X}$ is calculated for paramagnetic alignment of PAHs with $T_{\d}=60\K$ in the CNM with the typical magnetic field $B=10\mu$G.}
\label{fig:QX}
\end{figure}

Figure \ref{fig:QX} shows the different measures of grain alignment as functions of grain size. The effective Rayleigh reduction factor $R\cos^{2}\gamma$ is obtained from the best-fit model (Figure \ref{fig:Rfit}, upper). $Q_{X}$ is obtained from detailed calculations of grain alignment by paramagnetic relaxation in Hoang et al. (2013) for the cold neutral medium (CNM: $n_{\H}=30\cm^{-3}, T_{\gas}=100\K$) and a constant dust temperature $T_{\d}=60\K$, assuming a typical interstellar magnetic field $B=10\mu$G. $Q_{J}$ is calculated using Equation (\ref{eq:Ra}) with $f_{\rm corr}$ from Hoang et al. (2013). It can be seen that $Q_{J}$ is larger than $R$ by a factor of $3$.

\subsection{Polarized spinning dust emissivity}
The polarized emissivity and unpolarized emissivity of spinning dust emission can be given by
\bea
q_{\nu}&=&\int_{a_{l}}^{a_{u}} Q_{J}(a)\cos^{2}\gamma j_{\nu}(a) \frac{dn}{da} da ,\\
j_{\nu}&=&\int_{a_{l}}^{a_{u}} j_{\nu}(a) \frac{dn}{da}da, 
\ena
where $j_{\nu}(a)$ is the spinning dust emissivity at frequency $\nu$ from a grain of size $a$, and $a_{l}=3.56$\AA~ and $a_{u}=100$\AA. The frequency dependence of polarization of spinning dust emission is $p(\nu)=q_{\nu}/j_{\nu}$. We adopt the spinning dust model from \cite{2011ApJ...741...87H} for oblate spheroidal grains with ratio axis $r=2$ and a typical electric dipole moment $\beta=0.4$ Debye.

\begin{figure}
\includegraphics[width=0.5\textwidth]{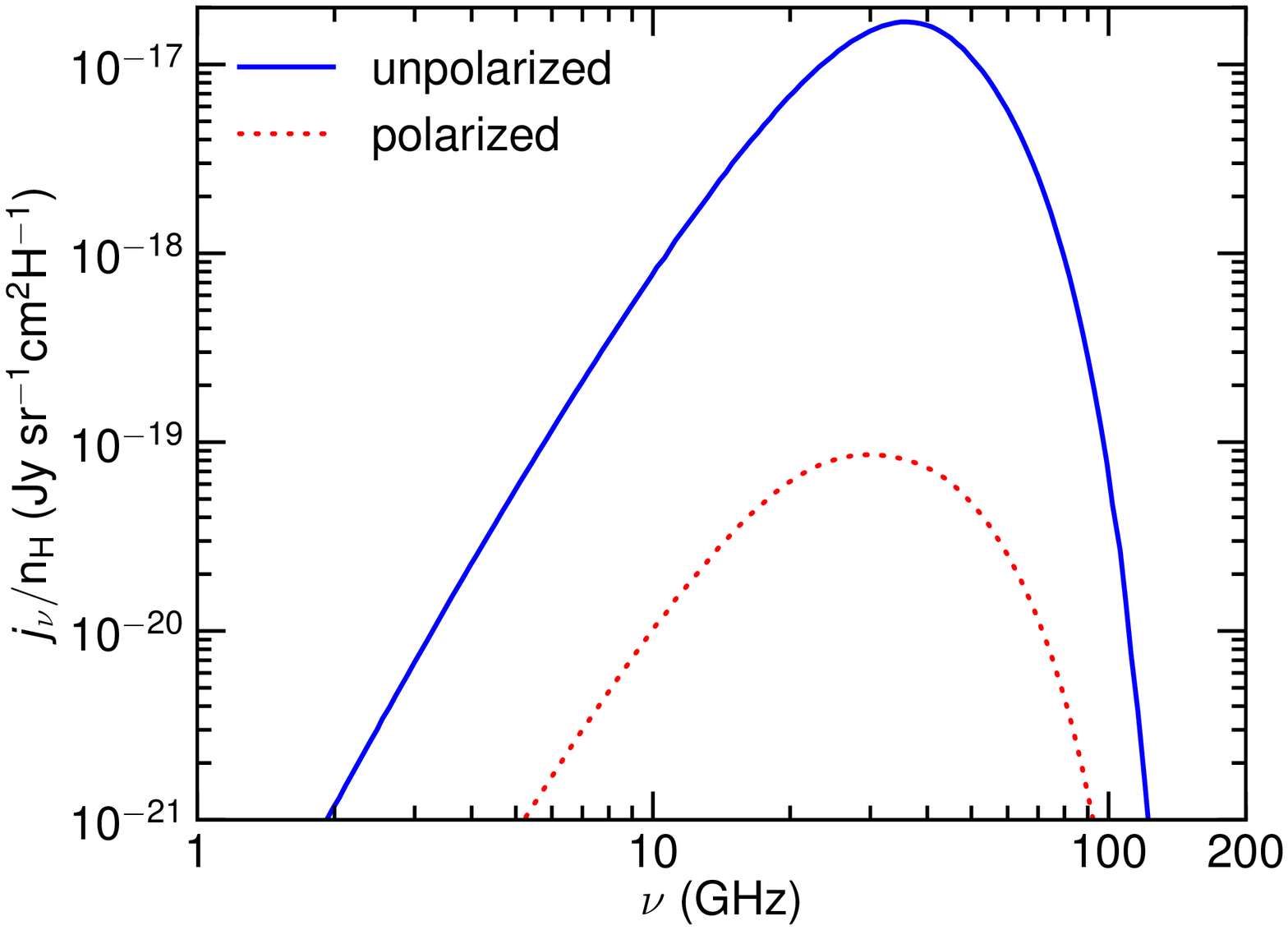}
\includegraphics[width=0.5\textwidth]{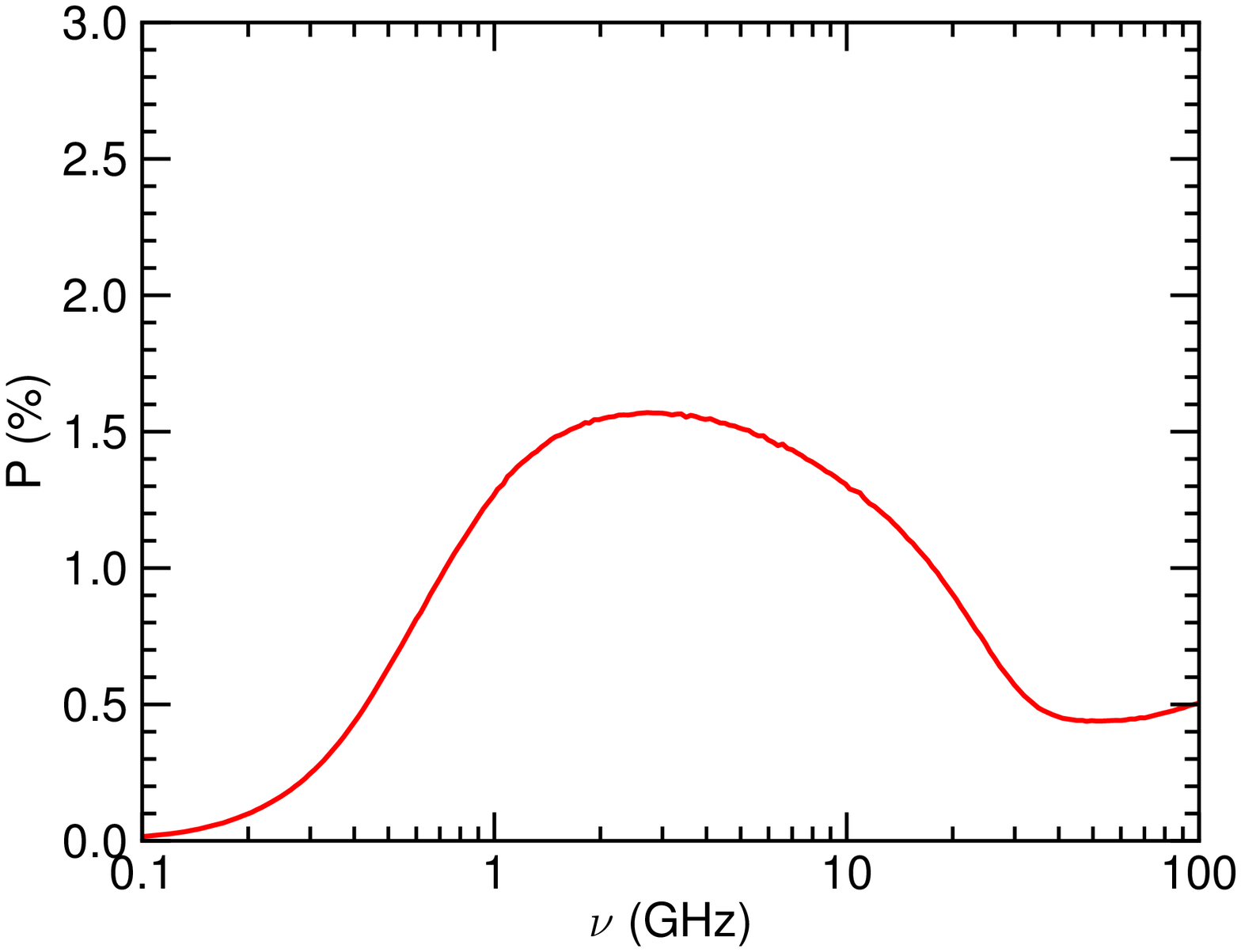}
\caption{Upper panel: Unpolarized (solid line) and polarized (dotted line) spinning dust emissivity from the diffuse ISM along the line of sight to HD 197770. Lower panel: Frequency dependence of degree of polarization of spinning dust emission.}
\label{fig:polspec}
\end{figure}

Figure \ref{fig:polspec} (upper panel) shows the total (unpolarized) and polarized spinning dust emissivity for the best-fit model. The degree of polarization is shown in Figure \ref{fig:polspec} (lower panel). It can be seen that the maximum degree of polarization is about $1.6\%$ and peaks at $\nu\approx 3\GHz$. The polarization is $\le 0.9\%$ for $\nu\ge 20 \GHz$. 

\section{Discussion}\label{sec:discus}
In the \Planck era, electric dipole emission from rapidly rotating PAHs has become an accepted component of diffuse galactic foreground emissions that contaminate to the CMB radiation. In light of more exciting \Planck results on CMB polarization, a pressing issue remains is to quantify the level of contamination by polarized spinning dust emission to the polarized CMB signal. The present paper seeks constraints on the polarization of spinning dust emission based on the models that best fit to observed polarization data of the stars having potential evidence of PAHs alignment.

\subsection{$2175$\AA~polarization bump and alignment of PAHs}

The $2175$\AA~polarization bump observed for two stars, HD197770 and HD147933-4, was discovered a long time ago (\citealt{1992ApJ...385L..53C}; \citealt{1993ApJ...403..722W}), and it was suggested that the polarization bump originates from aligned small graphite (see e.g. \citealt{1989IAUS..135..313D}). Employing the observational data for these stars, we obtain the grain size distributions and alignment functions that best reproduce both the observed extinction and polarization curves. 

For HD 197770, we found that a model with aligned silicate grains plus weakly aligned PAHs can successfully reproduce the $2175$\AA~polarization bump as well as the excess UV polarization. The effective degree of PAH alignment inferred for HD 197770 varies with the grain size and has peak $R\cos^{2}\gamma\approx 0.004$ at $a \approx 10$\AA. Although the degree of PAH alignment is rather small, due to the dominance of PAHs for the dust mass for $a<20$\AA, it is sufficient to reproduce the $2175$\AA~polarization bump. In fact, the low degree of alignment for PAHs is not unexpected from theoretical predictions based on resonance paramagnetic alignment, which was proposed by \cite{2000ApJ...536L..15L} and numerically studied in Hoang et al. 2013.
 
For HD 147933-4, on the other hand, we found no indication of alignment of PAHs. The model with only aligned silicate grains can account for the $2175$\AA~polarization bump; although the model overestimates the observed polarization for $\lambda^{-1}>5\mum^{-1}$ (see Figure \ref{fig:Rfit}).\footnote{More discussions on why it is challenging to obtain a satisfactory fit to the observed polarization data for the stars with large $\lambda_{\max}$ can be found in \cite{1995ApJ...444..293K}.}

Before understanding why there is such a difference in two stars, let us recall the significant difference in the optical and polarization properties of HD 197770 and HD 147933-4 (see Table \ref{tab:HD}). Indeed, the latter has a much higher ratio of visual-to-selective extinction $R_{V}$ and a much larger peak wavelength $\lambda_{\max}$. In addition, the polarization ratio $p$(2175\AA)/$p_{\max}$ is much lower in the later case. Qualitatively, $R_{V}$ reflects the average size of dust grains, whereas $\lambda_{\max}$ reflects the average size of {\it aligned} grains. Thus, compared to HD 197770, HD 147933-4 essentially has the larger average grain size and the larger average size of {\it aligned} grains. As a result, the UV polarization is dominated by the alignment of large grains. Moreover, since the large silicate grains have optical properties with oscillating features in the UV due to interference effects,\footnote{The oscillating feature is prominent for dust with the dielectric function having small imaginary part (weakly absorbing material).} they can induce the broad polarization feature seen beyond $\lambda^{-1}\approx 4.6\mum^{-1}$.

\subsection{Why is the 2175\AA~polarization bump not seen for most stars?}

The question now is that if the PAHs are potentially aligned, why we do not see the UV polarization feature for most of stars. To answer this question, let us consider the lines of sight of similar $R_{V}$ and $\lambda_{\max}$. Then the presence of $2175$\AA~polarization bump depends on the degree of alignment of PAHs in the ambient magnetic field. Although the alignment mechanism for PAHs is still not clear, the most promising mechanism is based on the resonance paramagnetic relaxation (LD00). Since the paramagnetic alignment is sensitive to the magnetic field and gas randomization, the variation of the magnetic field (both strength and direction) and gas density along different lines of sight can result in the absence of the UV polarization feature. 

In addition, the resonance paramagnetic alignment depends on spin-spin and spin-lattice relaxation within PAHs. If the relaxation is reduced, the expected degree of grain alignment is also reduced. The point of the relaxation in PAHs, as it was claimed in LD00 can be settled via laboratory studies.  

The possibility to identify the $2175$\AA~ polarization bump depends on both the alignment of PAHs and small silicate grains because the latter is responsible for the UV continuum polarization. If the alignment of small silicates is inefficient, then the bump can be detected due to high contrast. If the alignment of small silicate grains is considerable, the UV polarization produced by such grains tends to smooth out the bumpy polarization by PAHs, which makes the detection of 2175\AA~bump more difficult.

One interesting point with HD 197770 is that it has an excess UV polarization much lower than other stars with the same $\lambda_{\max}=0.51\mum$ (see \citealt{1995ApJ...445..947C}). Moreover, HD 197770 has an excess emission at $60\mum$, indicating that the radiation field intensity along this line of sight is higher than the averaged interstellar radiation field and the dust is hotter than the typical ISM. Since hotter, small grains tend to have the lower degree of alignment due to stronger thermal fluctuations within the grain, the UV continuum polarization is reduced accordingly, favoring the detection of the 2175\AA~polarization bump along this line of sight. 

Finally, the alignment degree of PAHs in the magnetic field in general is rather weak ($R\cos^{2}\gamma\le 0.004$) for which the resulting UV polarization excess would be small. As pointed out in \cite{1997ApJ...478..395W}, this feature might not be detected by low signal-to-noise observations.

\subsection{Constraint on the polarization of spinning dust emission}

We calculated the degree of polarization for spinning dust emission using the alignment function for the best-fit model to the HD 197770 data. We found that the upper limit for the polarization of spinning dust emission is $\sim 1.6\%$ at $\nu\approx 3\GHz$ and the degree of polarization declines rapidly to below $0.9\%$ for $\nu>20\GHz$. 

It is noted that our calculations assume that the UV polarization bump is produced by PAHs only (i.e., the contribution of large graphite grains is disregarded). In fact, if graphite grains can be aligned, then their contribution to the UV polarization bump would reduce our estimate for spinning dust polarization. Therefore, this constraint is indeed the upper limit for the spinning dust polarization. 

Our study assumed oblate spheroidal grains with axial ratio $r=2$ for both PAHs and silicate grains. It is obvious that for some grain shape with a smaller axial ratio, which corresponds to lower polarization efficiency $C_{\pol}$, the degree of alignment required to reproduce the $2175$\AA~ polarization bump is higher. As a result, the polarization of spinning dust is higher. However, ultrasmall grains or PAHs are expected to be planar, and the axial ratio appears to be $r \ge 2$ if the grains are approximated as oblate spheroid. Therefore, our constraint for spinning dust polarization obtained for $r=2$ plays as an upper limit.

Observationally, a number of studies indicate that anomalous microwave emission (AME) is weakly polarized (\citealt{2006ApJ...645L.141B}; \citealt{2009ApJ...697.1187M}; \citealt{LopezCaraballo:2011p508}). Using WMAP data, \cite{2011MNRAS.418..888M} found an upper limit for the polarization of AME between $1.4-2\%$ for $\nu=20-40\GHz$. \cite{RubinoMartin:2012ji} reviewed in great detail on observational constraints for the AME polarization, and an upper limit of $1\%$ for $\nu= 20-30\GHz$ is placed for the various environments. It appears that our upper limit on spinning dust polarization is consistent with the current observational data for the polarization of AME.

It is worth noting that in addition to spinning dust emission, magnetic dipole emission from dust was suggested to contribute some emission to the AME (\citealt{1999ApJ...512..740D}).  The degree of polarization of magnetic dipole emission may be considerable for the AME. However, an improved study in \cite{2013ApJ...765..159D} shows that the magnetic dipole emission is important for the AME at frequencies $\nu=20-40\GHz$ if a large fraction of the Fe is in metallic iron nanoparticles with extreme elongation.

\subsection{Implication for polarized far-infrared thermal dust emission}
\cite{Collaboration:2013ww} showed that the degree of polarization of thermal dust emission can reach a high level $P_{\rm em}>15\%$ at $\nu=353\GHz$. Let us estimate the degree of dust polarization at this frequency along the lines of sight toward two stars HD 197770 and HD 147933-4.

For simplicity, we assume that all grains have the same temperature. Thus, using the grain size distribution and alignment function from the model, we can obtain the degree of polarization of thermal dust emission as follows:
\bea
P_{\rm em}=\frac{I_{\lambda,\pol}}{I_{\lambda}}\approx \frac{\sigma_{\pol}(\lambda)}{\sigma_{\ext}(\lambda)},
\ena
where $I_{\lambda,\pol}$ is the polarized emission, $I_{\lambda}$ is the total emission, and $\sigma_{\ext}$ and $\sigma_{\pol}$ are present in Equations (\ref{eq:sigext}) and (\ref{eq:sigpol}).

At $\nu=353\GHz$ (i.e., $\lambda\approx 850\mum$), we found the degree of polarization $P_{\rm em}\approx 11\%$ and $14\%$ for HD 197770 and 147933-4, respectively. The higher degree of polarization for the HD 147933-4 star is obtained because this star has a higher fraction of large, aligned silicate grains which dominate far-infrared dust emission (see Figure \ref{fig:Rfit}). It can be seen that a high level ($\ge 15\%$) of dust polarization seen by \Planck can be expected from our model.

To study a potential correlation between the degree of polarization of far-infrared dust emission to that by dust extinction, we introduce a correlation parameter $r_{\pol}$, which is equal to the ratio of $P_{\rm em}$ at $\nu=353\GHz$ to $P_{\ext}/\tau$ at $\lambda=\lambda_{\max}$. We obtain $r_{\pol} \approx 5.2$ for HD 197770 and $r_{\pol}\approx 7.7$ for HD 147933-4.

\section{Summary}\label{sec:summ}
In the present paper, we obtain the constraint on the degree of polarization of spinning dust emission based on the observed extinction and polarization curves of starlight. Our principal results are summarized as follows:
\begin{itemize}

\item[1.] A model of grain size distribution and alignment function for interstellar grains that best fits to the observed extinction and polarization curve is obtained for two stars HD 197770 and HD147933-4 with the prominent polarization features at $2175$\AA.

\item[2.] We find that a small degree of alignment ($R\cos^{2}\gamma\approx 0.004$) of PAHs included to the aligned silicate grains is sufficient to reproduce the $2175$\AA polarization feature for HD 197770. For HD 147933-4, we find no indication of alignment for PAHs and the alignment of silicate grains can account for the feature.

\item[3.] We calculate the polarized spinning dust emissivity using the alignment function from the best-fit model. We show that the degree of polarization for spinning dust emission has a peak of $\approx 1.6\%$ at $\nu\approx 3\GHz$ and rapidly declines to below $0.9\%$ for $\nu>20 \GHz$.

\item[4.] The degree of polarization for thermal dust emission at $353\GHz$ is estimated to be $P_{\rm em}\approx 11\%$ for the line of sight to the HD 197770 star and $P_{\rm em}\approx 14\%$ for the line of sight to the HD 147933-4 star.
\end{itemize}

\acknowledgments
We thank the anonymous referee for valuable comments and suggestions that significantly improved our paper. T.H. thanks Brandon Hensley for useful discussions. A.L. acknowledges the support of the NASA Grant NNX11AD32G and the NSF grant AST-1109469, as well as  the Vilas Associate Award and the support of the NSF Center for Magnetic Self-Organization. A part of the work was performed during A.L. stay in the stimulating atmosphere of the International Institute of Physics (Brazil).

%--------------adding references-----------------------------------
%\bibliographystyle{/Users/thiemhoang/Dropbox/Papers2/apj}
% or other styles: mcbride,plain, abbrv, acm, alpha, apalike, apj
%\bibliography{/Users/thiemhoang/Dropbox/Papers2/cites_paperApJ,/Users/thiemhoang/Dropbox/Papers2/cites_Books}
\bibliography{ms.bbl}

\end{document}